\documentclass[12pt,leqno]{article}

\usepackage{amsmath,amssymb}
\usepackage{latexsym}
\usepackage{graphicx}
\usepackage{bm}

\def\R{\mathbb{R}}

\begin{document}
Proc. Amer. Math. Soc. 141, N2,(2013), 515-521.

\baselineskip=16pt

\title{Symmetry problem}

\author{ A. G. Ramm$\dag$\footnotemark[1] \\
\\
$\dag$Mathematics Department, Kansas State University,\\
Manhattan, KS 66506-2602, USA
}

\renewcommand{\thefootnote}{\fnsymbol{footnote}}
\footnotetext[1]{Email: ramm@math.ksu.edu} 

\date{}
\maketitle
\begin{abstract}
\baselineskip=16pt
 A novel approach to an old symmetry problem is developed.
A proof is 
given for the following symmetry problem, studied earlier:
if $\Delta u=1$ in $D\subset \R^3$, $u=0$ on $S$, the boundary of $D$,
and $u_N=const$ on $S$, then $S$ is a sphere. 
It is assumed that $S$ is a Lipschitz surface homeomorphic to a sphere.
This result has been proved
in different ways by various authors. Our proof is based on a simple 
new idea.

\end{abstract}



\section{Introduction}

Symmetry problems are of interest both theoretically and in applications.
A well-known, and still unsolved, symmetry problem is the Pompeiu problem 
(see \cite{R363}, \cite{R382}, and the references therein). In modern 
formulation this problem consists of 
proving the following conjecture: 

{\it If $D\subset \mathbb{R}^n,\, n\ge 2,$
is a domain homeomorphic to a ball, and the boundary $S$ of $D$ is 
smooth ($S\in C^{1,\lambda},\, \lambda>0,$ is sufficient), and if the 
problem
\begin{equation}
 (\nabla^2 + k^2)u = 0\quad \text{in}\quad D,\qquad u\big{|}_S =c\neq 0, 
\quad u_N\big{|}_S=0, \quad k^2 = const>0,
\end{equation}
where $c$ is a constant, has a solution, then $S$ is a 
sphere.}

A similar problem ({\it Schiffer's conjecture}) is also unsolved
(see also \cite{CH}): 

{\it If the 
problem
\begin{equation}
 (\nabla^2 + k^2)u = 0\quad \text{in}\quad D,\qquad u\big{|}_S = 0,
\quad u_N\big{|}_S=c\not=0,
 \quad k^2 = const>0
\end{equation}
has a solution, then $S$ is a sphere.}

{\bf Standing assumptions}: 

{\it In this paper we assume  that $D\subset 
\mathbb{R}^3$
is a bounded domain
homeomorphic to a ball, with a sufficiently smooth boundary $S$ ($S$ is
Lipschitz suffices).}

We use the following notation: $D'=\mathbb{R}^3\setminus D$, $B_R = \{ x: 
|x|\le R \}$, $B_R\supset D$,
$\mathcal{H}$ is
the set of all harmonic functions
in $B_R$, $R>0$ is an arbitrary large number, such that the ball $B_R$
contains $D$, 
$|D|$ and $|S|$ are
the volume of $D$ and the surface area of $S$, respectively.
 
In 
\cite{R512} it 
is proved that if
\begin{equation}
\int_D \frac{dy}{4\pi |x-y|} = \frac{c}{|x|},\qquad \forall x\in B'_R, 
\quad c = const>0,
\end{equation}
then $D$ is a ball. The proof in \cite{R512} is based on an idea similar
to the one we are using in this paper.

In \cite{R468} a symmetry problem of interest in elasticity theory is
studied by  A.D.Alexandrov's method of a moving plane (\cite{A}), used
also in \cite{S}. The result in \cite{S}, which is formulated below
in Theorem 1, was proved in \cite{W} by a method, different from the one
given in \cite{S}, and discussed also in \cite{Am}. The argument
in \cite{Am} remained unclear to the author.

In \cite{R556} another symmetry problem of potential theory was studied.
 
Our goal is to give a new proof of Theorem 1. The result
of Theorem 1 was obtained 
in \cite{S} for $\R^n$, $n\ge 2$.

{\bf Theorem 1.} {\it If $D\subset \R^3$ is a bounded domain, homeomorphic 
to a ball,  $S$ is its Lipschitz boundary, and the  problem:
\begin{equation}
\label{eq4}
\Delta u = 1\quad \text{in}\quad D,\qquad u\big{|}_S = 0,
\quad u_N\big{|} = c:=\frac {|D|}{|S|} >0
\end{equation}
has a solution, then $S$ is a sphere.}

This result is equivalent to the following result:

{\it If 
\begin{equation}
\label{eq5}
\int_D h(x)dx = c\int_S h(s)ds,
\qquad \forall h\in \mathcal{H},\qquad c:=\frac {|D|} {|S|},
\end{equation}
then $S$ is a sphere.}

The equivalence of \eqref{eq4} and \eqref{eq5} can be proved as follows. 

Suppose
\eqref{eq4} holds. Multiply \eqref{eq4} by an arbitrary $h\in \mathcal{H}$, 
integrate by parts and get
\begin{equation}
\label{eq6}
\int_D h(x)dx = c \int_S h(s)ds.
\end{equation}
If $h=1$ in \eqref{eq6}, then one gets $c=\frac{|D|}{|S|}$, 
so \eqref{eq6} is identical to \eqref{eq5}.

Suppose \eqref{eq5} holds. Then \eqref{eq6} holds. Let $v$ solve the 
problem $\Delta v = 1$
in $D$, $v\big{|}_S=0$. This $v$ exists and is unique. Using
\eqref{eq6}, the equation $\Delta h = 0$ in $D$, and the Green's 
formula, one gets
\begin{equation}
c \int_S h(s)ds = \int_D h(x)dx = \int_D h(x)\Delta v dx = \int_S 
h(s)v_N ds.
\end{equation} 
Thus,
\begin{equation}
\label{eq8}
\int_S h(s)[v_N - c ]ds = 0,\qquad \forall h\in \mathcal{H}.
\end{equation}
We will need the following lemma:

{\bf Lemma A.} {\it
The set of restrictions on $S$ of all harmonic functions in $D$ is dense 
in $L^2(S)$.}
 
{\bf Proof of Lemma A.} We give a proof for the convenience of the reader.
The proof is borrowed from \cite{R512}. Suppose that $g\in L^2(S)$, 
and $\int_Sg(s)h(s)ds=0\quad \forall h\in \mathcal{H}.$ Since $(4\pi 
|x-y|)^{-1}$
is in $\mathcal{H}$ if $y\in D'$, one gets 
$$w(y):=\int_Sg(s)(4\pi|s-y|)^{-1}ds=0\quad \forall y\in D'.$$  
Thus, a single layer potential $w$, with $L^2$ density $g$, vanishes
in $D'$, and, by continuity, on $S$. Since $w$ is a harmonic function in 
$D$ vanishing on $S$, it follows that $w=0$ in $D$. By the jump formula 
for the normal derivative of the single-layer potential across a Lipschitz 
boundary, one gets $g=0$. \hfill $\Box$

Thus, \eqref{eq8} implies $v_N\big{|}_S = c$. Therefore, \eqref{eq4} 
holds.

A result, related to equation \eqref{eq5}, was studied in \cite{K}
for a two-dimensional problem. The arguments in \cite{K} were not
quite clear to the author. 

Our main result is a new proof of Theorem 1. The proof is 
simple,  and the method of the proof is new. This method can be used in 
other problems (see \cite{R556}, \cite{R382},\cite{R512}, \cite{R470}).

\section{Proofs}
{\it Proof of Theorem 1.}
We denote by $D'$ the complement of $D$ in $\R^3$, by 
$S^2$ the unit sphere, by $[a,b]$ the cross product of two vectors, 
by $g=g(\phi)$ the rotation about an axis, directed along a vector  
$\alpha\in S^2$, by the angle $\phi$, and note that if $h(x)$
is a harmonic function in any ball $B_R$, containing $D$,
then $h(gx)$ is also a harmonic function in $B_R$.

Take $h=h(g(\phi)x)$ in \eqref{eq5}, differentiate with respect to $\phi$ 
and then set $\phi=0$. This yields:
$$\int_D \nabla h(x)\cdot [\alpha, x]dx=c\int_S\nabla h(s)\cdot [\alpha, 
s]ds.$$
Using the divergence theorem, one rewrites this as:
$$\alpha \cdot \int_S [s, N] h(s)ds=\alpha \cdot \int_S [s,c\nabla 
h(s)]ds.$$
Since $\alpha\in S^2$ is arbitrary, one gets
\begin{equation}
\label{eq9}
\int_S [s, N] h(s)ds=\int_S [s,c\nabla h(s)]ds, \qquad \forall h\in 
\mathcal{H},
\end{equation}
where $N=N_s$ is a unit normal to
$S$ at the point $s\in S$, pointing into $D'$. 

Let $y\in B_R'$ be an arbitrary point, and $h(x)= \frac 
{1}{|x-y|}\in \mathcal{H}$, where $x\in B_R$. Then
equation \eqref{eq9} implies that
\begin{equation}
\label{eq10} 
v(y):=\int_S \frac {[s, N]ds}{|s-y|}=c[\nabla \int_S\frac {ds}{|s-y|}, 
y],\qquad \forall y\in B_R',
\end{equation}
because
\begin{equation}
\label{eq11}c\int_S[s,\nabla_s\frac {1}{|s-y|}]ds=c\int_S[\frac 
s{|s-y|^3}, y]ds=
c[\nabla_y \int_S\frac {ds}{|s-y|}, y].
\end{equation}
Relation \eqref{eq11} actually holds for all $y\in D'$, because 
of the analyticity of its left and right sides in $D'$. 
Let $$w(y):=\int_S|s-y|^{-1}ds.$$ 


Denote $y^0:=y/|y|$. 
It is known (see, e.g., \cite{BE}) that
\begin{equation}
\label{eq12'}|y-s|^{-1}=\sum_{n=0}^\infty \sum_{m=-n}^n \frac 
{4\pi}{2n+1}Y_{nm}(y^0)
\overline{Y_{nm}(s^0)}|s|^n |y|^{-(n+1)},\qquad |y|>|s|,
\end{equation}
where the overline stands for the complex conjugate,
$y^0$ is the unit vector characterized by the angles $\theta, \phi$
in spherical coordinates, $Y_{nm}$ are normalized spherical harmonics:
$$Y_{nm}(y^0)=Y_{nm}(\theta, \phi)=\gamma_{nm}P_{n,|m|}(\cos\theta)e^{im\phi},
\qquad -n\le m\le n,$$
$\gamma_{nm}=[\frac{(2n+1)(n-m)!}{4\pi (n+m)!}]^{1/2}$ are normalizing 
constants: 
$$(Y_{nm}(y^0),Y_{pq}(y^0))_{L^2(S^2)}=\delta_{np}\delta_{mq},$$
and 
$$P_{n,|m|}(\cos\theta)=(\sin\theta)^{|m|}
(\frac{d}{d\cos\theta})^{|m|}P_n(\cos\theta)$$
are the associated Legendre functions, where $P_n(\cos\theta)$ are the 
Legendre
polynomials.

If $z=\cos\theta$, then 
$$P_{n,m}(z)=(z^2-1)^{m/2}(\frac{d}{dz})^{m}P_n(z), \quad 
m=1,2,...,$$
$$P_n(z)=(2^n n!)^{-1} (\frac{d}{dz})^{n}(z^2-1)^{n}, \quad P_0(z)=1$$
(see \cite{BE}). The definitions of $P_{n,m}(z)$ in various
books can differ by a factor $(-1)^m$.

Using formula \eqref{eq12'}, one obtains
\begin{equation}
\label{eq12}
w(y)=\sum_{n=0}^\infty \frac {4 \pi}{2n+1}\sum_{m=-n}^n 
Y_{nm}(y^0)|y|^{-(n+1)} c_{nm},\qquad 
c_{nm}:=\int_S|s|^{n}\overline{Y_{nm}(s^0)}ds.  
\end{equation}

Substitute this in \eqref{eq10}, equate the terms in front of 
$|y|^{-(n+1)}$, and define vectors
\begin{equation}
\label{eq13} 
a_{nm}:=\int_S[s,N] |s|^{n}\overline{Y_{nm}(s^0)}ds
\end{equation}
to obtain
\begin{equation}
\label{eq14}
\sum_{m=-n}^nY_{nm}(y^0)a_{nm}=\sum_{m=-n}^ncc_{nm}[e_\theta\partial_\theta
Y_{nm}(y^0) +e_\phi (\sin \theta )^{-1}\partial_\phi Y_{nm}(y^0), e_r],
\end{equation}
where $e_\theta$,$e_\phi$, and $e_r$ are orthogonal unit vectors of the 
spherical coordinate system, $[e_\phi,e_r]$ is the cross product, 
$[e_\phi,  e_r]=e_\theta,$ $[e_\theta,  e_r]=-e_\phi$,
$y=ry^0$, $r=|y|$,  $y^0=(\sin \theta \cos \phi, 
\sin \theta \sin \phi, \cos \theta)$, 
$\partial_\theta=\frac \partial{\partial 
\theta}$. 

Therefore, formula \eqref{eq14}
can be rewritten as
\begin{equation}
\label{eq15}
\sum_{m=-n}^nY_{nm}(y^0)a_{nm}=\sum_{m=-n}^ncc_{nm}\Big{(}-e_\phi 
\partial_\theta Y_{nm}(y^0) +e_\theta (\sin \theta )^{-1}\partial_\phi 
Y_{nm}(y^0)\Big{)}.
\end{equation}

{\it From \eqref{eq15} we want to derive that} 
\begin{equation}
\label{eq16}
a_{nm}=0, \qquad n\ge 0, \, -n\le m\le n.
\end{equation}
If \eqref{eq16} is established, then it follows
from \eqref{eq13} and from the completeness in $L^2(S)$
of the system $\{|s|^nY_{nm}(s^0)\}_{n\ge 0, -n\le m\le n}$  that 
$[s,N]=0$ on $S$,
and this implies that $S$ is a sphere, as follows from Lemma 1
formulated and proved below. Consequently, 
Theorem 1 is proved as soon as relations \eqref{eq16} are established.
The completeness of the system $\{|s|^nY_{nm}(s^0)\}_{n\ge 0, -n\le m\le 
n}$ in $L^2(S)$ follows from Lemma B:

The functions $|x|^nY_{nm}(x^0)$, $n\ge 0$, $-n\le m \le n,$  are 
harmonic in any ball, centered at the origin.

{\bf Lemma B.} {\it The set of
restrictions of the above functions to any Lipschitz surface homeomorphic 
to a sphere is complete in $L^2(S)$.}

{\bf Proof of Lemma B.} The proof is given for completeness. It is similar 
to the proof of Lemma A. Suppose that $g\in L^2(S)$ and 
$$\int_S g(s)|s|^nY_{nm}(s^0)ds=0,\quad \forall n\ge 0, |m|\le n.$$
This and \eqref{eq12'} imply that
$$\int_S g(s)(4\pi |s-y|)^{-1}ds=0 \qquad \forall y\in D',$$
and the argument, given in the proof of Lemma A, yields the desired
conclusion $g=0$. \hfill $\Box$

Vector $a_{nm}$ is written in Cartesian basis $\{e_j\}_{1\le j \le 3}$ as
$$a_{nm}=\sum_{j=1}^3 a_{nm,j}e_j.$$ 
The relation between the
components $F_1, F_2, F_3,$ of a vector $F$ in Cartesian coordinates 
and its components $F_r,F_\theta, F_\phi,$ in spherical coordinates
can be found, e.g.,  in \cite{KK}, Section 6.5:
$$F_1=F_r \sin\theta\cos\phi +F_\theta \cos\theta \cos\phi -F_\phi \sin 
\phi,$$
$$F_2=F_r \sin\theta\sin\phi +F_\theta \cos\theta \sin\phi +F_\phi \cos
\phi,$$
$$F_3=F_r \cos\theta  -F_\theta \sin\theta.
$$
Using these relations one derives
from \eqref{eq15} the following formulas:
\begin{equation}
\label{eq17}
\sum_{m=-n}^na_{nm,1}Y_{nm}(y^0)=\sum_{m=-n}^n 
cc_{nm}\Big(\partial_\theta Y_{nm}(y^0)\sin\phi +
\partial_\phi Y_{nm}(y^0)\cot\theta\cos \phi \Big), 
\end{equation}
\begin{equation}
\label{eq18}
\sum_{m=-n}^na_{nm,2}Y_{nm}(y^0)=\sum_{m=-n}^n 
cc_{nm}\left(-\partial_\theta Y_{nm}(y^0)\cos\phi 
+\partial_\phi Y_{nm}(y^0)\cot \theta
\sin \phi\right),
\end{equation}
\begin{equation}
\label{eq19}
\sum_{m=-n}^na_{nm,3}Y_{nm}(y^0)=-\sum_{m=-n}^n 
cc_{nm}\partial_\phi Y_{nm}(y^0).
\end{equation}

{\it From formulas  \eqref{eq17}-\eqref{eq19} one derives \eqref{eq16}}.

If $n=0$,  then $a_{00}=0$, as the following calculation shows: 
$$a_{00}=\frac 1{(4\pi)^{1/2}}\int_S [s,N]ds=-\frac 
1{(4\pi)^{1/2}}\int_D[\nabla, x]dx=0.$$

If $n>0$, then multiply equation \eqref{eq19} by $e^{-im\phi}$, integrate 
with respect to $\phi$ over $[0,2\pi]$, write $P_{n,m}$ for 
$P_{n,m}(\cos\theta)$, and obtain 
\begin{equation}
\label{eq20}
a_{nm,3}P_{n,m}=-cc_{n,m} im P_{n,m},\qquad c_{n,m}:=c_{nm}.
\end{equation}
One concludes that $a_{n0,3}=0$ and $a_{nm,3}=-im cc_{n,m}$.

If one derives from \eqref{eq17}-\eqref{eq18} that $c_{n,m}=0$,
then equation \eqref{eq16} follows,  and Theorem 1 is proved.

From \eqref{eq17} and \eqref{eq18} one derives analogs of \eqref{eq20}:
\begin{equation}
\label{eq21}\begin{split} 
2ia_{nm,1}\gamma_{nm}P_{n,m}&=cc_{n,m-1}\gamma_{n,m-1}\left(\partial_\theta
P_{n,m-1}-(m-1)\cot\theta P_{n,m-1}\right)\\
&-cc_{n,m+1}\gamma_{n,m+1}\left(\partial_\theta P_{n,m+1}+(m+1)\cot\theta 
P_{n,m+1}\right),
\end{split}
\end{equation}
\begin{equation}
\label{eq22}\begin{split}
2a_{nm,2}\gamma_{nm}P_{n,m}&=cc_{n,m-1}\gamma_{n,m-1}\left(-\partial_\theta
P_{n,m-1}+(m-1)\cot\theta P_{n,m-1}\right)\\
&-cc_{n,m+1}\gamma_{n,m+1}\left(\partial_\theta P_{n,m+1}+(m+1)\cot\theta
P_{n,m+1}\right).
\end{split}
\end{equation}
Let us take $\theta\to 0$ in the above equations. It is known
(see \cite{BE}, Section 3.9.2, formula (4)) that
\begin{equation}
\label{eq23}
P_{n,m}(z)\sim b(n,m)(z-1)^{m/2}, \quad z\to 1, \qquad 
b(n,m):=\frac{(n+m)!}{2^{m/2}m!(n-m)!}.
\end{equation}
Equation \eqref{eq21} can be considered as a linear
combination 
\begin{equation}
\label{eq24}
\sum_{j=1}^3 A_jf_j(z)=0, 
\end{equation}
where the $A_j$ are constants:
$$A_1=2ia_{nm,1}\gamma_{nm},\quad A_2=-cc_{n,m-1}\gamma_{n,m-1},\quad
A_3=cc_{n,m+1}\gamma_{n,m+1},$$ and
$$f_1(z)=P_{n,m}(z),$$ 
$$f_2(z)= -(1-z^2)^{1/2} P'_{n,m-1}(z)-(m-1)\frac {z}{(1-z^2)^{1/2}} 
P_{n,m-1}(z),$$ 
$$f_3(z)=-(1-z^2)^{1/2} P'_{n,m+1}(z)-(m+1)\frac 
{z}{(1-z^2)^{1/2}}P_{n,m+1}(z),\quad z=\cos\theta.$$
If the system of functions $\{f_j(z)\}_{j=1}^3$ is linearly independent on 
the interval $[-1,1]$, then all $A_j=0$ in \eqref{eq24}, that is,
$A_1=0$, $A_2=0$, and $A_3=0$. 
This implies that 
$$a_{nm,1}=c_{n,m}=0, \quad -n\le m\le n.$$ 
The quantities $a_{nm,2}$ and $a_{nm,3}$ are proportional to 
$c_{n,m}$. Since $c_{n,m}=0$, it follows that 
$$a_{nm,2}=a_{nm,3}=0,\quad -n\le m\le n,$$   
and  Theorem 1 is proved.

{\it Thus, to complete the proof of Theorem 1 it is sufficient to verify 
the linear 
independence of the system of functions $\{f_j(z)\}_{j=1}^3$ on the 
interval $z\in [-1,1]$.}
 
 From formula \eqref{eq23} it follows that these functions
have the following main terms of 
their asymptotics as $z\to 1$: 
$$f_1(z)\sim B_1(z-1)^{m/2},\quad f_2(z)\sim B_2 \frac 
{(z-1)^{(m+1)/2}}{(1-z^2)^{1/2}},\quad f_3(z)\sim B_3 
\frac{(z-1)^{(m+3)/2}}{(1-z^2)^{1/2}},$$ where 
the constants $B_j\neq 0$, $1\le j\le 3$, depend on $n,m$.
The  linear independence 
of the system $\{f_j(z)\}_{j=1}^3$ holds because the system   
$$\{(z-1)^{m/2},\quad \frac {(z-1)^{(m+1)/2}}{(1-z^2)^{1/2}},\quad 
\frac{(z-1)^{(m+3)/2}}{(1-z^2)^{1/2}}\}$$ 
is linearly independent.
The linear independence of this system holds if the system
$$\{1,\quad (1+z)^{-0.5}, \quad (z-1)(1+z)^{-0.5}\}$$ is linearly 
independent on the interval $[-1,1]$.
The linear independence of this system on the interval $[-1,1]$ is 
obvious.

Theorem 1 is proved. \hfill $\Box$

{\bf Lemma 1.} {\it If $S$ is a $C^2-$smooth closed surface
and $[s,N_s]=0$ on $S$, then $S$ is a sphere.}

{\it Proof of Lemma 1.} Let $s=r(u,v)$ be a parametric equation of $S$. 
Then the vectors $r_u$ and $r_v$ are linearly independent and 
$N_s$ is directed along the vector $[r_u, r_v]$. Thus, the assumption
$[s,N_s]=0$ on $S$ implies that 
$$[r,[r_u, r_v]]=r_u(r,r_v)-r_v(r,r_u)=0.$$
Since the vectors $r_u$ and $r_v$ are linearly  independent, it follows 
that  $(r,r_v)=(r,r_u)=0$. Thus, $(r,r)=R^2$, where $R^2$ is a constant.
This means that $S$ is a sphere. Lemma 1 is proved. \hfill $\Box$


\newpage

\begin{thebibliography}{99}

\baselineskip=16pt

\bibitem{A} A. D. Alexandrov, {\it A characteristic property of a sphere}, 
Ann. di Matem., 58, (1962), 303-315.

\bibitem{Am} T. Amdeberhan, {\it Two symmetry problems in potential 
theory}, Electronic Journ. of Diff. Eqs, 43, (2001), 1-5.  

\bibitem{BE} H. Bateman, A. Erdelyi, {\it Higher transcendental 
functions},
Vol. 1, McGraw-Hill, New York, 1953.

\bibitem{CH} T. Chatelain and A. Henrot, Some results about Schiffer's 
conjectures,  Inverse Problems,  15, (1999), 647-658. 

\bibitem{R556} N. S. Hoang and A. G. Ramm, {\it Symmetry problems 2}, 
Annal. Polon. Math., 96, N1, (2009), 61-64.

\bibitem{KK} G. Korn and T. Korn, {\it Mathematical Handbook for
Scientists and Engineers,} 
McGraw-Hill, New York, 1968. 

\bibitem{K} A. A. Kosmodem'yanskii, {\it A converse of the mean value 
theorem for harmonic functions}, Russ. Math. Surveys, 36, N5, (1981), 
159-160.

\bibitem{R190} A. G. Ramm, {\it Scattering by obstacles}, D.Reidel, 
Dordrecht, 1986. 

\bibitem{R363} A. G. Ramm, {\it The Pompeiu problem}, Applicable Analysis, 
64, N1-2, 
(1997), 19-26.

\bibitem{R382} A. G. Ramm, 
{\it Necessary and sufficient condition for a domain, which fails to have 
Pompeiu property, to be a ball}, Journ. of
Inverse and Ill-Posed Probl., 6, N2, (1998), 165-171.

\bibitem{R470} A. G. Ramm, {\it Inverse Problems}, Springer, New York, 
2005.

\bibitem{R512} A. G. Ramm,
{\it A symmetry problem}, Ann. Polon. Math., 92, (2007), 49-54.

\bibitem{R468} A. G. Ramm and  E. Shifrin,
{\it Symmetry problems in the elasticity theory problem for plane cracks 
of normal rapture}, Journ. of Appl. Math. and Mech.,
69, (2005), 127-134.



\bibitem{S} J. Serrin, {\it A symmetry problem in potential theory}, Arch. 
Rat. Mech. Anal.,
43, (1971), 304-318.

\bibitem{W} H. Weinberger, {\it Remark on the preceding paper of Serrin},
Arch. Rat. Mech. Anal., 43, (1971), 319-320.


\end{thebibliography}
\end{document}